\documentclass[twocolumn,showpacs,preprintnumbers,amsmath,amssymb]{revtex4-1}

\usepackage{CJK}

\usepackage{graphicx}
\usepackage{dcolumn}
\usepackage{bm}
\usepackage{tabularx}

\begin{document}
\begin{CJK*}{GBK}{song}

\title{Determination of astrophysical $^{12}$N($p$,\,$\gamma$)$^{13}$O reaction rate
from the $^2$H($^{12}$N,\,$^{13}$O)$n$ reaction and its astrophysical implications}

\author{B. Guo} \author{J. Su} \author{Z. H. Li} \author{Y. B.
Wang} \author{S. Q. Yan} \author{Y. J. Li}
\author{N. C. Shu} \author{Y. L. Han}
\author{X. X. Bai}
\author{Y. S. Chen}
\author{W. P. Liu}\thanks{Corresponding author: wpliu@ciae.ac.cn}
\affiliation{China Institute of Atomic Energy, P.O. Box 275(10),
Beijing 102413, China}
\author{H. Yamaguchi} \author{D. N. Binh} \author{T. Hashimoto} \author{S. Hayakawa} \author{D. Kahl}
\author{S. Kubono} \affiliation{Center for Nuclear Study
(CNS), the University of Tokyo, Wako Branch, 2-1 Hirosawa, Wako,
Saitama 351-0198, Japan}
\author{J. J. He}
\author{J. Hu} \author{S. W. Xu}
\affiliation{Institute of Modern Physics, Chinese Academy of
Sciences (CAS), Lanzhou 730000, China} \author{N. Iwasa}
\author{N. Kume} \affiliation{Department of
Physics, Tohoku University, 6-6 Aoba, Sendai, Miyagi 980-8578,
Japan} \author{Z. H. Li}\affiliation{School of Physics and
State Key Laboratory of Nuclear Physics and Technology, Peking
University, Beijing 100871, China}

\date{\today}

\begin{abstract}
The evolution of massive stars with very low-metallicities depends
critically on the amount of CNO nuclides which they produce. The
$^{12}$N($p$,\,$\gamma$)$^{13}$O reaction is an important branching
point in the rap-processes, which are believed to be alternative
paths to the slow 3$\alpha$ process for producing CNO seed nuclei
and thus could change the fate of massive stars. In the present
work, the angular distribution of the $^2$H($^{12}$N,\,$^{13}$O)$n$
proton transfer reaction at $E_{\mathrm{c.m.}}$ = 8.4 MeV has been
measured for the first time. Based on the Johnson-Soper approach,
the square of the asymptotic normalization coefficient (ANC) for the
virtual decay of $^{13}$O$_\mathrm{g.s.}$ $\rightarrow$ $^{12}$N +
$p$ was extracted to be 3.92 $\pm$ 1.47 fm$^{-1}$ from the measured
angular distribution and utilized to compute the direct component in
the $^{12}$N($p$,\,$\gamma$)$^{13}$O reaction. The direct
astrophysical S-factor at zero energy was then found to be 0.39
$\pm$ 0.15 keV b. By considering the direct capture into the ground
state of $^{13}$O, the resonant capture via the first excited state
of $^{13}$O and their interference, we determined the total
astrophysical S-factors and rates of the
$^{12}$N($p$,\,$\gamma$)$^{13}$O reaction. The new rate is two
orders of magnitude slower than that from the REACLIB compilation.
Our reaction network calculations with the present rate imply that
$^{12}$N($p,\,\gamma$)$^{13}$O will only compete successfully with
the $\beta^+$ decay of $^{12}$N at higher ($\sim$two orders of
magnitude) densities than initially predicted.
\end{abstract}

\maketitle

\section{Introduction}

The first generation of stars formed at the end of the cosmic dark
ages, which marked the key transition from a homogeneous and simple
universe to a highly structured and complex one \cite{bro04}. The
first stars of zero metallicity are so-called Population III that
formed before Population I in galatic disks and Population II in
galatic halos \cite{cay86,car87}. The most fundamental question
about Population III stars is how massive they typically were since
the mass of stars dominates their fundamental properties such as
lifetimes, structures and evolutions. Recent numerical simulations
of the collapse and fragmentation of primordial gas clouds indicate
that these stars are predominantly very massive with masses larger
than hundreds of $M_\odot$ (see Ref. \cite{bro04} and references
therein).

A classic question on the evolution of supermassive stars is whether
they contributed any significant material to later generations of
stars by supernova explosions which ended the lives of Population
III stars. In 1986, Fuller, Woosley and Weaver \cite{ful86} studied
the evolution of non-rotating supermassive stars with a hydrodynamic
code KEPLER. They concluded that these stars will collapse into
black holes without experiencing a supernova explosion. This is
because the triple alpha process (3$\alpha \rightarrow $ $^{12}$C)
does not produce sufficient amounts of CNO seed nuclei so that the
hot CNO cycle and $rp$-process are unable to generate the nuclear
energy enough to explode the stars. In 1989, Wiescher, G\"{o}rres,
Graff et al. \cite{wie89} suggested the rap-processes as alternative
paths which would permit these stars to bypass the 3$\alpha$ process
and to yield the CNO material. The reactions involved in the
rap-processes are listed as below:

$\textrm{rap-I}:
~^3\textrm{He}(\alpha,\,\gamma)^7\textrm{Be}(p,\,\gamma)^8\textrm{B}(p,\,\gamma)^9\textrm{C}(\alpha,\,p)^{12}\textrm{N}(p,\,\gamma)$\\
\indent~~~~~~~~~~$^{13}\textrm{O}(\beta^+\nu)^{13}\textrm{N}(p,\,\gamma)^{14}\textrm{O}$\\
\indent $\textrm{rap-II}:
~^3\textrm{He}(\alpha,\,\gamma)^7\textrm{Be}(\alpha,\,\gamma)^{11}\textrm{C}(p,\,\gamma)^{12}\textrm{N}(p,\,\gamma)$\\
\indent~~~~~~~~~~~$^{13}\textrm{O}(\beta^+\nu)^{13}\textrm{N}(p,\,\gamma)^{14}\textrm{O}$\\
\indent $\textrm{rap-III}:
~^3\textrm{He}(\alpha,\,\gamma)^7\textrm{Be}(\alpha,\,\gamma)^{11}\textrm{C}(p,\,\gamma)^{12}\textrm{N}(\beta^+\nu)$\\
\indent~~~~~~~~~~~~$^{12}\textrm{C}(p,\,\gamma)^{13}\textrm{N}(p,\,\gamma)^{14}\textrm{O}$\\
\indent $\textrm{rap-IV}:
~^3\textrm{He}(\alpha,\,\gamma)^7\textrm{Be}(\alpha,\,\gamma)^{11}\textrm{C}(\alpha,\,p)^{14}\textrm{N}(p,\,\gamma)^{15}\textrm{O}$.\\
It is crucial to determine the rates of the key reactions in the
rap-processes in order to study if they play any significant role in
the evolution of supermassive stars by producing CNO material. $^{12}$N($p$,\,$\gamma$)$^{13}$O
is an important reaction in the rap-I and rap-II processes.

Due to the low Q-value (1.516 MeV) of the
$^{12}$N($p$,\,$\gamma$)$^{13}$O reaction, its stellar reaction rate
is dominated by the direct capture into the ground state in
$^{13}$O. In addition, the resonant capture via the first excited
state in $^{13}$O could play an important role for determining the
reaction rates. In 1989, Wiescher et al. \cite{wie89} derived the
direct astrophysical S-factor at zero energy, S(0), to be $\sim$40
keV b based on a shell model calculation. In 2006, Li \cite{liz06a}
extracted the direct S(0) factor to be 0.31 keV b by using the
spectroscopic factor from the shell model calculation of Ref.
\cite{war06}, where the proton-removal cross section of $^{13}$O on
a Si target was well reproduced. It should be noted that there is a
discrepancy of two orders of magnitude between the above two values
of the direct S(0) factor. In 2009, Banu, Al-Abdullah, Fu et al.
\cite{ban09} derived the asymptotic normalization coefficient (ANC)
for the virtual decay of $^{13}$O$_\mathrm{g.s.}$ $\rightarrow$
$^{12}$N + $p$ from the measurement of the
$^{14}$N($^{12}$N,\,$^{13}$O)$^{13}$C angular distribution and then
calculated the direct S(0) factor to be 0.33 $\pm$ 0.04 keV b, which
is consistent with that in Ref. \cite{liz06a}. As for the resonant
capture component, the resonant parameters of the first excited
state in $^{13}$O have been studied through a thick target technique
\cite{ter03,sko07} and $R$-matrix method \cite{sko07,ban09}. In
1989, Wiescher et al. \cite{wie89} derived the radiative width to be
$\Gamma_\gamma$ = 24 meV with one order of magnitude uncertainty
based on a Weisskopf estimate of the transition strength. In 2007,
Skorodumov, Rogachev, Boutachkov et al. \cite{sko07} measured the
excitation function of the resonant elastic scattering of $^{12}$N +
$p$ and extracted the spin and parity to be $J^\pi$ = 1/2$^+$ for
the first excited state in $^{13}$O via an $R$-matrix analysis. In
addition, the excitation energy and the proton width were determined
to be 2.69 $\pm$ 0.05 MeV and 0.45 $\pm$ 0.10 MeV, respectively. In
2009, Banu et al. \cite{ban09} derived a radiative width
$\Gamma_\gamma$ = 0.95 eV by using the experimental ANC, based on
the $R$-matrix approach.

This work aims at determining the astrophysical S-factors and rates
of the $^{12}$N($p,\,\gamma$)$^{13}$O reaction through the ANC
approach based on an independent proton transfer reaction. Here, the
angular distribution of the $^{12}$N($d$,\,$n$)$^{13}$O reaction
leading to the ground state in $^{13}$O is measured in inverse
kinematics, and used to extract the ANC for the virtual decay of
$^{13}$O$_\mathrm{g.s.}$ $\rightarrow$ $^{12}$N + $p$ through the
Johnson-Soper adiabatic approximation \cite{wal76}. The ($d$,\,$n$)
transfer system has been successfully applied to the study of some
proton radiative capture reactions, such as
$^{7}$Be($p,\,\gamma$)$^{8}$B \cite{liu96,das06},
$^{8}$B($p,\,\gamma$)$^{9}$C \cite{bea01},
$^{11}$C($p,\,\gamma$)$^{12}$N \cite{liu03}, and
$^{13}$N($p,\,\gamma$)$^{14}$O \cite{liz06b}. The astrophysical
S-factors and rates for the direct capture in the
$^{12}$N($p,\,\gamma$)$^{13}$O reaction are then calculated by using
the measured ANC. Finally, we obtain the total S-factors and rates
by taking into account the direct capture, the resonant capture and
their interference, and study the temperature-density conditions at
which the $^{12}$N($p,\,\gamma$)$^{13}$O reaction takes place.

\section{Measurement of the $^2$H($^{12}$N,\,$^{13}$O)$n$ angular distribution}

The experiment was performed with the CNS low energy in-flight
Radio-Isotope Beam (CRIB) separator \cite{kub02,yan05} in the RIKEN
RI Beam Factory (RIBF). A $^{10}$B primary beam with an energy of 82
MeV was extracted from the AVF cyclotron. The primary beam impinged
on a $^3$He gas target with a pressure of 360 Torr and a temperature
of 90 K; the target gas was confined in a small chamber with a
length of 80 mm \cite{yam08}. The front and rear windows of the gas
chamber are Havar foils, each in a thickness of 2.5 $\mu$m. The
secondary $^{12}$N ions with an energy of 70 MeV were produced
through the $^3$He($^{10}$B,\,$^{12}$N)$n$ reaction and then
selected by the CRIB separator, which mainly consists of two
magnetic dipoles and a velocity filter (Wien filter).

\begin{figure}[htbp]
\begin{center}
\resizebox{0.45\textwidth}{!}{
  \includegraphics{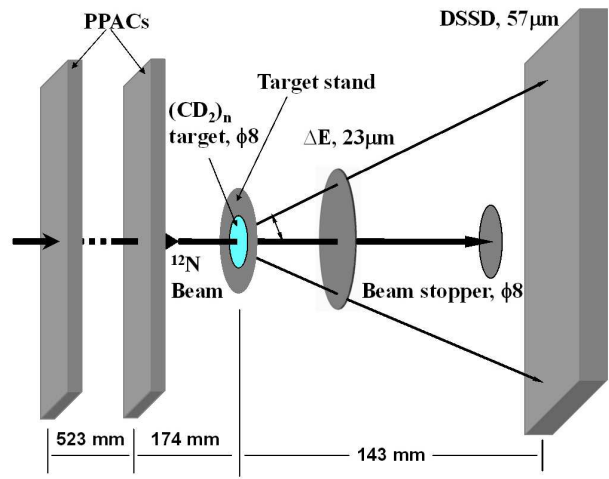}}
\caption{(Color online) Schematic layout of the experimental setup
at the secondary reaction chamber (namely F3 chamber in Ref.
\cite{yan05}) of CRIB for the
$^2$H($^{12}$N,\,$^{13}$O$_{\mathrm{g.s.}}$)$n$ reaction.}
\label{fig:1}
\end{center}
\end{figure}

\begin{figure}[htbp]
\begin{center}
\resizebox{0.48\textwidth}{!}{
  \includegraphics{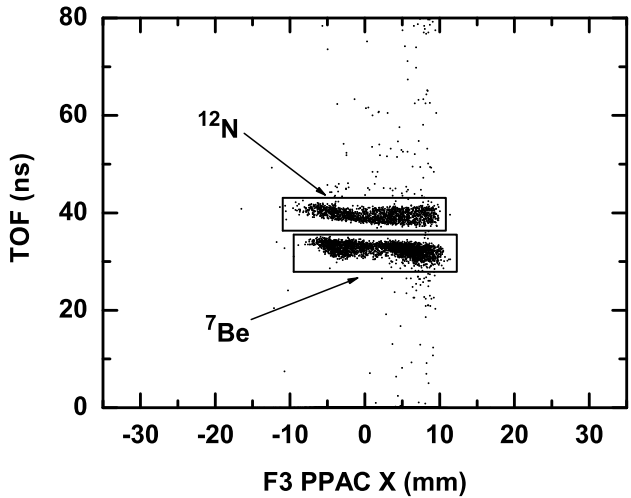}}
\caption{Two-dimensional histogram of TOF vs. horizontal position
(X) on the upstream PPAC in the F3 chamber for the particle
identification of the cocktail beam.} \label{fig:2}
\end{center}
\end{figure}

A schematic layout of the experimental setup at the secondary
reaction chamber (namely F3 chamber, see Ref. \cite{yan05} for
details) of CRIB separator is shown in Fig. \ref{fig:1}. The
cocktail beam which included $^{12}$N was measured event-by-event
using two parallel plate avalanche counters (PPACs) \cite{kum01}; in
this way, we determined the particle identification, precise timing
information, and could extrapolate the physical trajectory of each
ion in real space. In Fig. \ref{fig:2} we display the histogram of
time of flight (TOF) vs. horizontal position (X) on the upstream
PPAC in the F3 chamber for the particle identification of the
cocktail beam. The main contaminants are $^{7}$Be ions with the
similar magnetic rigidities and velocities to the $^{12}$N ions of
interest. After the two PPACs, the $^{12}$N secondary beam bombarded
a deuterated polyethylene (CD$_2$) film with a thickness of 1.5
mg/cm$^2$ to study the $^2$H($^{12}$N,\,$^{13}$O)$n$ reaction. A
carbon film with a thickness of 1.5 mg/cm$^2$ was utilized to
evaluate the background contribution from the carbon nuclei in the
(CD$_2$) target. The target stand with a diameter of 8 mm also
served as a beam collimator. The typical purity and intensity of the
$^{12}$N ions on target were approximately 30\% and 500 pps after
the collimator, respectively.

The $^{13}$O reaction products were detected and identified with a
telescope consisting of a 23 $\mu$m silicon detector ($\Delta E$)
and a 57 $\mu$m double-sided silicon strip detector (DSSD). In order
to determine the energy of $^{12}$N ions after they pass through two
PPACs, a silicon detector with a thickness of 1500 $\mu$m was placed
between the downstream PPAC and the (CD$_2$) target, and removed
after measuring the beam energy. The energy calibration of the
detectors was carried out by combining the use of $\alpha$-source
and the magnetic rigidity parameters of $^{10}$B and $^{12}$N ions.
The energy loss of the $^{12}$N beam in the whole target was determined from
the energy difference measured with and without the target.
The $^{12}$N beam energy in the middle of the (CD$_2$) target was
derived to be 59 MeV from the energy loss calculation by the program
LISE++ \cite{tar08}, which was calibrated by the experimental energy loss
in the whole target. In addition, a beam stopper (close to the DSSD) with a
diameter of 8 mm was used to block un-reacted beam particles in
order to reduce radiation damage to the DSSD.

\begin{figure}[htbp]
\begin{center}
\resizebox{0.5\textwidth}{!}{
  \includegraphics{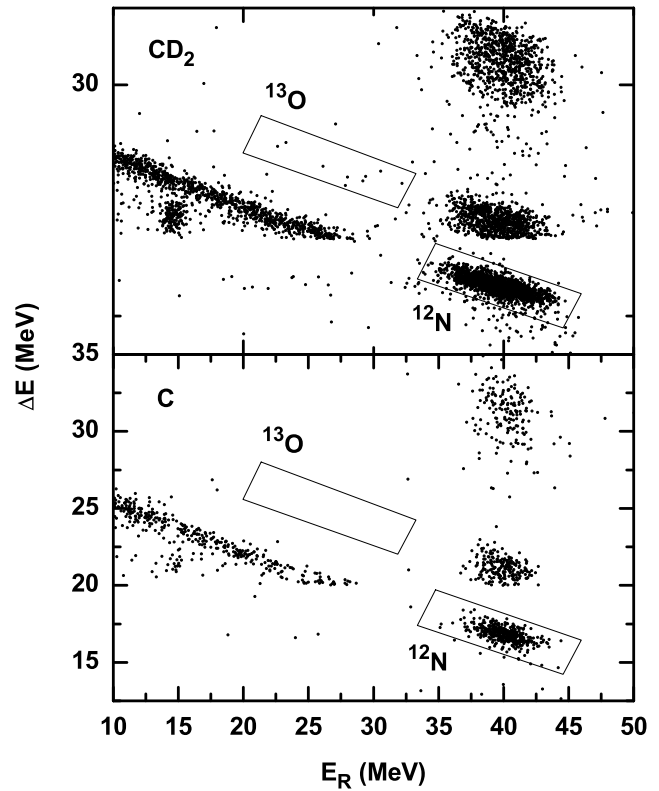}}
\caption{Two-dimensional histogram of energy loss $\Delta E$ vs.
residual energy $E_\mathrm{R}$ for the events in the angular range
of 3$^\circ <$ $\theta_\mathrm{c.m.} <$ 4$^\circ$. The top and
bottom panels display the spectra for the measurement of the
deuterated polyethylene (CD$_2$) and carbon (C) targets,
respectively. The two-dimensional cuts for the $^{13}$O events from
the $^2$H($^{12}$N,\,$^{13}$O)$n$ reaction were determined with a MC
simulation. See text for details.} \label{fig:3}
\end{center}
\end{figure}

\begin{figure}[htbp]
\begin{center}
\resizebox{0.48\textwidth}{!}{
  \includegraphics{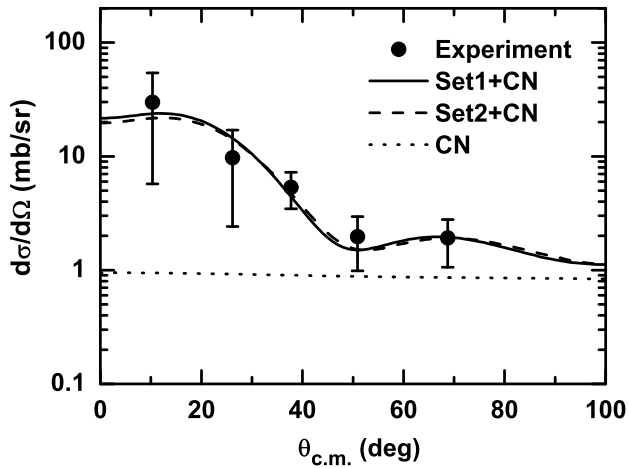}}
\caption{Measured angular distribution of
$^2$H($^{12}$N,\,$^{13}$O$_{\mathrm{g.s.}}$)$n$ at $E_\mathrm{c.m.}$
= 8.4 MeV, together with the theoretical calculations on the
direct-process contribution using two sets of optical potentials
(Set1 and Set2) and the compound-nucleus (CN) contribution. See the
text in the section \ref{analysis} for the details on the
theoretical calculations.} \label{fig:4}
\end{center}
\end{figure}

The emission angles of reaction products were determined by
combining the position information from the DSSD and the two PPACs.
As an example, Fig. \ref{fig:3} shows a two-dimensional histogram of
energy loss ($\Delta E$) vs. residual energy ($E_\mathrm{R}$) for
the events in the angular range of 3$^\circ <$ $\theta_\mathrm{c.m.}
<$ 4$^\circ$. For the sake of saving CPU time in dealing with the
experimental data, all the events below $\Delta E$ = 20 MeV were
scaled down by a factor of 100, and the $^{13}$O events were not
affected. The two-dimensional cuts of the $^{13}$O events from the
$^2$H($^{12}$N,\,$^{13}$O)$n$ reaction were determined with a Monte
Carlo (MC) simulation, which took into account the kinematics,
geometrical factor, the energy diffusion of the $^{12}$N beam, the
angular straggling, and the energy straggling in the two PPACs, the
secondary target and the $\Delta E$ detector. This simulation was calibrated
by using the $^{12}$N elastic scattering on the target. Such a calibration
approach has been successfully used to study the $^2$H($^8$Li,\,$^9$Li)$^1$H
reaction \cite{guo05}. The $^{13}$O events
are clearly observed in the two-dimensional cut for the (CD$_2$)
measurement, while no relevant events are observed in this cut for
the background measurement. The $^{7}$Be contaminants don't affect
the identification of the $^{13}$O events since these ions and their
products are far from the $^{13}$O region in the spectra of $\Delta
E$ vs. $E_\mathrm{R}$ and have significantly different energies from
the $^{13}$O events. The effects of the pileup of $^{7}$Be with
$^{12}$N can be estimated and subtracted through the background
measurement. In addition, the detection efficiency correction from
the beam stopper was also computed via the MC simulation also by
considering the effects mentioned above. The resulting detection
efficiencies range from 66\% to 100\% for different detection
regions in the DSSD. After the beam normalization and background
subtraction, the angular distribution of the
$^2$H($^{12}$N,\,$^{13}$O$_\mathrm{g.s.}$)$n$ reaction in the center
of mass frame was obtained and is shown in Fig. \ref{fig:4}.

\section{Analysis of the $^2$H($^{12}$N,\,$^{13}$O)$n$ angular
distribution}\label{analysis}

For a peripheral transfer reaction, the ANC can be derived by the
comparison of the experimental angular distribution with theoretical
calculations,
\begin{eqnarray}
\left( {\frac{d\sigma}{d\Omega}} \right)_{\mathrm{exp}} =
\sum_{j_{i}j_{f}}(C_{l_{i}j_{i}}^{d})^{2}(C_{l_{f}j_{f}}^{^{13}\mathrm{O}})^{2}R_{l_{i}j_{i}l_{f}j_{f}},
\label{eq1}
\end{eqnarray}
where
\begin{eqnarray}
R_{l_{i}j_{i}l_{f}j_{f}}={\sigma^{th}_{l_{i}j_{i}l_{f}j_{f}} \over
(b_{l_{i}j_{i}}^{d})^{2}(b_{l_{f}j_{f}}^{^{13}\mathrm{O}})^{2}}.
\label{eq2}
\end{eqnarray}
$\left( {\frac{d\sigma}{d\Omega}} \right)_{\mathrm{exp}}$ and
$\sigma^{th}_{l_{i}j_{i}l_{f}j_{f}}$ are the experimental and
theoretical differential cross sections, respectively.
$C_{l_{f}j_{f}}^{^{13}\mathrm{O}}$, $C_{l_{i}j_{i}}^{d}$ and
$b_{l_{f}j_{f}}^{^{13}\mathrm{O}}$, $b_{l_{i}j_{i}}^{d}$ represent
the nuclear ANCs and the corresponding single particle ANCs for the
virtual decays of $^{13}$O$_\mathrm{g.s.}$ $\rightarrow$ $^{12}$N +
$p$ and $d$ $\rightarrow$ $p$ + $n$, respectively. $l_{i}$, $j_{i}$
and $l_{f}$, $j_{f}$ denote the orbital and total angular momenta of
the transferred proton in the initial and final nuclei $d$ and
$^{13}$O, respectively. $R_{l_{i}j_{i}l_{f}j_{f}}$ is model
independent in the case of a peripheral transfer reaction;
therefore, the extraction of the ANC is insensitive to the geometric
parameters (radius $r_{0}$ and diffuseness $a$) of the bound state
potential.

\begin{table}
\caption{\label{tab1} Optical potential parameters used in the
calculation, where $V$ and $W$ are in MeV, $r$ and $a$ in fm.}
\begin{center}
\begin{tabular}{ccccc}
\hline
\hline Set No. & 1 \cite{var91} && 2 \cite{kon03}& \\
\hline Channel & Entrance & Exit & Entrance & Exit \\
\hline
$V_{r}$ &97.03 &53.44 & 99.84 &55.44\\
$r_{0r}$ &1.152  &1.154 &1.127 &1.131\\
$a_{r}$ &0.722 &0.69 &0.708&0.676\\
$W$ &1.73 & &0.86& 0.77\\
$r_{w}$ &1.693  & &1.693& 1.131\\
$a_{w}$ &0.716 & &0.711& 0.676\\
$W_{s}$ &14.01 &9.61 &14.2&9.54\\
$r_{0s}$ &1.147 &1.147&1.306&1.306\\
$a_{s}$ &0.716 &0.716&0.56&0.56\\
$V_{so}$ &5.9 &5.9&5.65&5.65\\
$r_{0so}$ &0.816 &0.83&0.903&0.907\\
$a_{so}$ &0.661 &0.63&0.622&0.622\\
$r_{0c}$ &1.25 & &1.25& \\
\hline\hline
\end{tabular}
\end{center}
\end{table}

\begin{figure}[htbp]
\begin{center}
\resizebox{0.48\textwidth}{!}{
  \includegraphics{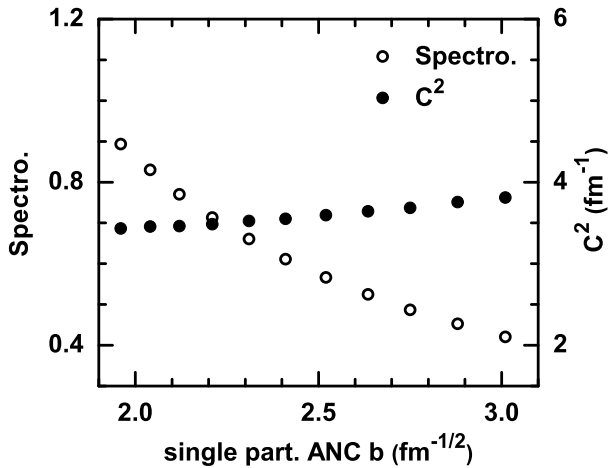}}
\caption{Variation of the spectroscopic factor (Spectro.) and the
square of the ANC ($C^2$) for the virtual decay of
$^{13}$O$_\mathrm{g.s.}$ $\rightarrow$ $^{12}$N + $p$ as a function
of the single particle ANC $b$.} \label{fig:5}
\end{center}
\end{figure}

In this work, the code FRESCO \cite{tho88} was used to analyze the
experimental angular distribution. In order to include the breakup
effects of deuterons in the entrance channel, the angular
distribution was calculated within the Johnson-Soper adiabatic
approximation to the neutron, proton, and target three-body system
\cite{wal76}. In the present calculation, the optical potentials of
nucleon-target were taken from Refs. \cite{var91,kon03}, which have
been successfully applied to the study of some of the reactions on
light nuclei \cite{tsa05,liu04,guo07}. The theoretical angular
distributions of the direct process were calculated with these two
sets of optical potentials, as shown in Fig. \ref{fig:4}. The
employed optical potential parameters are listed in Table
\ref{tab1}. In addition, the UNF code \cite{zha02} was used to
evaluate the compound-nucleus (CN) contribution in the
$^2$H($^{12}$N,\,$^{13}$O$_\mathrm{g.s.}$)$n$ reaction, as indicated
by the dotted line in Fig. \ref{fig:4}. The single-particle bound
state wave functions were calculated with conventional Woods-Saxon
potentials whose depths were adjusted to reproduce the binding
energies of the proton in the ground states of the deuteron ($E_b$ =
2.225 MeV) and $^{13}$O ($E_b$ = 1.516 MeV). To verify if the
transfer reaction is peripheral, the ANCs and the spectroscopic
factors were computed by changing the geometric parameters of
Woods-Saxon potential for single-particle bound state, using one set
of the optical potential, as displayed in Fig. \ref{fig:5}. One sees
that the spectroscopic factors depend significantly on the selection
of the geometric parameters, while the ANC is nearly constant,
indicating that the $^2$H($^{12}$N,\,$^{13}$O$_\mathrm{g.s.}$)$n$
reaction at the present energy is dominated by the peripheral
process.

The spins and parities of $^{12}$N$_\mathrm{g.s.}$ and
$^{13}$O$_\mathrm{g.s.}$ are $1^+$ and $3/2^-$, respectively.
Therefore, the $^2$H($^{12}$N,\,$^{13}$O$_\mathrm{g.s.}$)$n$ cross
section could include two contributions from the proton transfers to
1$p_{3/2}$ and 1$p_{1/2}$ orbits in $^{13}$O. The ratio of
1$p_{3/2}$:1$p_{1/2}$
[$(C_{p_{3/2}}^{^{13}\mathrm{O}})^{2}/(C_{p_{1/2}}^{^{13}\mathrm{O}})^{2}$]
was derived to be 0.16 based on a shell model calculation
\cite{war06}. $C_{d}^{2}$ was taken to be 0.76 fm$^{-1}$ from Ref.
\cite{blo77}. After the subtraction of the CN contribution, the
first three data points at forward angles were used to derive the
ANC by the comparison of the experimental data with the theoretical
calculations. For one set of optical potential, three ANCs can be
obtained by using three data points, and their weighted mean was
then taken as the final value. The square of the ANCs for the
1$p_{1/2}$ and 1$p_{3/2}$ orbits were extracted to be
$(C_{p_{1/2}}^{^{13}\mathrm{O}})^{2}$ = 3.38 $\pm$ 1.27 fm$^{-1}$
and $(C_{p_{3/2}}^{^{13}\mathrm{O}})^{2}$ = 0.54 $\pm$ 0.20
fm$^{-1}$, respectively. Consequently, the square of total ANC was
$(C_\mathrm{tot}^{^{13}\mathrm{O}})^{2}$ = 3.92 $\pm$ 1.47
fm$^{-1}$. The error resulted from the measurement (36\%) and the
uncertainty of the optical potentials (12\%). This result is in
agreement with the value of $(C_{p_{1/2}}^{^{13}\mathrm{O}})^{2}$ =
2.53 $\pm$ 0.30 fm$^{-1}$ obtained from the
$^{14}$N($^{12}$N,\,$^{13}$O)$^{13}$C reaction \cite{ban09}.

\section{Astrophysical rate of the $^{12}$N($p,\,\gamma$)$^{13}$O reaction and its astrophysical implications}

\begin{figure}[htbp]
\begin{center}
\resizebox{0.48\textwidth}{!}{
  \includegraphics{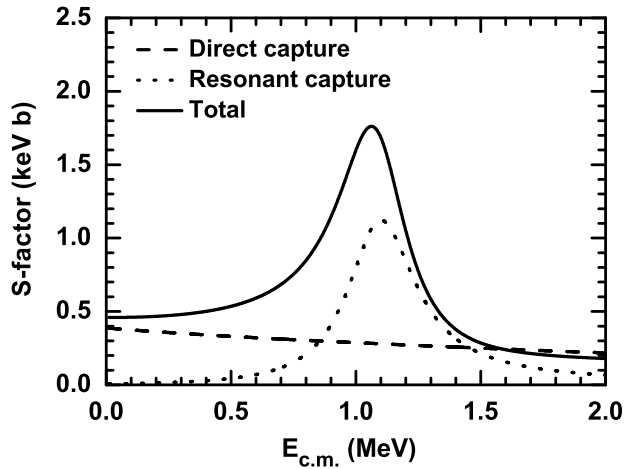}}
\caption{Astrophysical S-factors of the
$^{12}$N($p,\,\gamma$)$^{13}$O reaction as a function of
$E_\mathrm{c.m.}$. The dashed and dotted curves represent the direct
and resonant components, respectively. The solid curve denotes the
total S-factor. See text for details.} \label{fig:6}
\end{center}
\end{figure}

The ANC, which defines the amplitude of the tail of the radial
overlap function, determines the overall normalization of the direct
astrophysical S-factors \cite{muk01}. In the present work, the
direct capture cross sections and astrophysical S-factors were
computed based on the measured ANC by using the RADCAP code
\cite{ber03}, which is a potential model tool for direct capture
reactions. The resulting direct astrophysical S-factors as a
function of $E_\mathrm{c.m.}$ are displayed in Fig. \ref{fig:6}, as
indicated by the dashed line. The S-factor at zero energy was then
found to be $S(0)$ = 0.39 $\pm$ 0.15 keV b, which agrees with the
values in Refs. \cite{liz06a,ban09}.

The astrophysical S-factors of the resonant capture can be obtained
by using Breit-Wigner formula. In the present calculation, the
resonant parameters ($J^\pi$ = 1/2$^+$, $E_x$ = 2.69 $\pm$ 0.05 MeV,
$\Gamma_p$ = 0.45 $\pm$ 0.10 MeV \cite{sko07}, and $\Gamma_\gamma$ =
0.95 $\pm$ 0.24 eV \cite{ban09}) were adopted. In Fig. \ref{fig:6},
we display the resulting S-factors for the resonant capture, as
indicated by the dotted line.

Interference effects will occur only in the case that the resonant
and direct amplitudes have the same channel spin $I$ and the same
incoming orbital angular momentum \cite{ban09,muk99}. The direct
capture amplitude for the $^{12}$N($p,\,\gamma$)$^{13}$O reaction is
given by the sum of $I$ = 1/2 and 3/2 components. Since the channel
spin for the first resonance is 1/2, only the first component in the
direct capture interferes with the resonant amplitude. Therefore,
the total S-factors were calculated with
\begin{equation}
\label{eq3} S_{t}(E)=S_{d}(E)+S_{r}(E) \pm
2[S^{1/2}_{d}(E)S_{r}(E)]^{1/2}\cos(\delta),
\end{equation}
where $S_{d}(E)$, $S_{r}(E)$ and $S^{1/2}_{d}(E)$ denote the
astrophysical S-factors for the direct capture, the resonant
capture, and the $I$=1/2 component in the direct capture,
respectively. $\delta$ is the resonance phase shift, which can be
given by
\begin{equation}
\label{eq4} \delta=\arctan\Big{[}{\Gamma_{p}(E) \over
2(E-E_{R})}\Big{]}.
\end{equation}
Here, $\Gamma_p(E)=\Gamma_p {P_{l_i}(E) \over P_{l_i}(E_R)}$, where
$P_{l_i}(E)$ is the penetration factor. The ratio of the $I$=1/2
amplitude to the total amplitude in the direct capture was derived
to be 2/3 using the RADCAP code. Generally, the sign of the
interference in Equation \ref{eq3} has to be determined
experimentally. However, it is also possible to infer this sign via
an $R$-matrix method. Recently, Banu et al. \cite{ban09} found the
constructive interference below the resonance and the destructive
one above it using an $R$-matrix approach. Based on this
interference pattern, the present total S-factors were then
obtained, as shown in Fig. \ref{fig:6}. In addition, we estimated
the uncertainty of the total S-factors by taking into account the
errors of the present ANC for the ground state in $^{13}$O and the
employed resonant parameters for the first excited state in
$^{13}$O.

The astrophysical $^{12}$N($p,\,\gamma$)$^{13}$O reaction rates
(cm$^3$s$^{-1}$mol$^{-1}$) were then calculated with
\cite{rol88,ili07}
\begin{eqnarray}
\label{eq5}%
N_A \langle\sigma v\rangle= N_A\Big({8 \over \pi\mu}\Big)^{1/2}{1
\over (kT)^{3/2}}\int^{\infty}_0 S(E)\nonumber\\
\exp\Big[-\Big({E_{G} \over E}\Big)^{1/2}-{E \over kT}\Big]dE,
\end{eqnarray}
where the Gamow energy $E_{G}=0.978Z^{2}_{1}Z^{2}_{2}\mu$ MeV, and
$N_{A}$ is Avogadro's number.

\begin{figure}[htbp]
\begin{center}
\resizebox{0.48\textwidth}{!}{
  \includegraphics{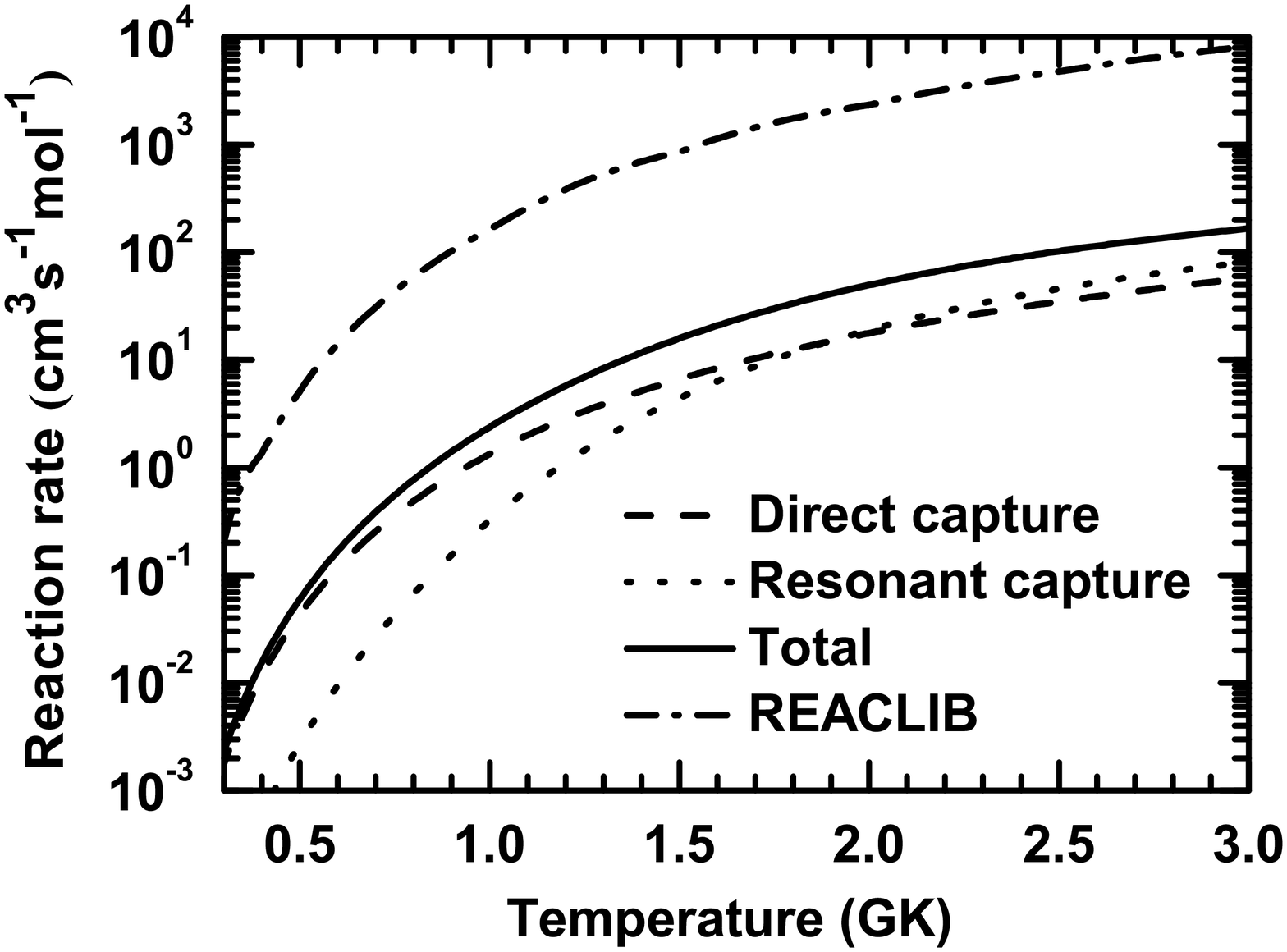}}
\caption{Astrophysical rates of the $^{12}$N($p,\,\gamma$)$^{13}$O
reaction as a function of temperature. The dashed and dotted curves
represent the present rates of the direct and resonant captures,
respectively. The solid curve represents the total rates in the
present work, while the dash-dotted curve denotes the REACLIB
compilation \cite{wie89}. See text for details.} \label{fig:7}
\end{center}
\end{figure}

In Fig. \ref{fig:7} we display the resulting reaction rates as a
function of temperature, together with the REACLIB compilation
\cite{wie89}. There is a discrepancy of up to a factor of $\sim$100
between these two reaction rates for the temperatures below $T_{9}$
= 3 ($T_9$ is a temperature in unit of 10$^9$ K). In addition, our
total rates are in good agreement with those in Fig. 9 of Ref.
\cite{ban09} since the similar contribution of the direct capture
was found and the same resonant parameters were used in both works.
From Fig. \ref{fig:7} one also sees that the direct capture
dominates the $^{12}$N($p,\,\gamma$)$^{13}$O reaction for the
temperatures below $T_{9}$ = 1.5.

\begin{table}
\caption{\label{tab2} The coefficients $a_i$ in Eq. \ref{eq6} for
the central value, lower limit and upper limit of the
$^{12}$N($p,\,\gamma$)$^{13}$O reaction rate. The fitting errors are
all less than 6\% at $T_9$ = 0.01-10.}
\begin{center}
\begin{tabular}{p{1cm}p{2cm}p{2cm}p{2cm}}
\hline
\hline $a_i$ & Central value & Upper limit& Lower limit\\
\hline
$a_{1}$ & -5.91219& -3.77150& -4.48818\\
$a_{2}$ & 0.0400148 & 0.0224401& 0.0253469\\
$a_{3}$ & -22.2259& -20.3542& -20.5376\\
$a_{4}$ & 32.2983& 28.2893& 28.6331\\
$a_{5}$ & -3.56352& -3.29024& -3.36174\\
$a_{6}$ & 0.261761& 0.243627& 0.252225\\
$a_{7}$ & -10.1115& -8.37104& -8.49396\\
$a_{8}$ & -10.0637& 4.00598& -7.05380\\
$a_{9}$ & -0.553380& 0.190095& -0.291547\\
$a_{10}$ & -20.7778& -31.6274& -20.5490\\
$a_{11}$ & 31.5635& 31.0575& 27.9040\\
$a_{12}$ & -3.64300& -6.70434& -3.31062\\
$a_{13}$ & 0.248400& -0.399390& 0.244458\\
$a_{14}$ & -12.2610& -13.7936& -10.4292\\
\hline \hline
\end{tabular}
\end{center}
\end{table}

We fitted the new rates with an expression used in the astrophysical
reaction rate library REACLIB \cite{thi87,rau01}. The total reaction
rates were fitted as
\begin{eqnarray}
\label{eq6}%
N_A \langle\sigma v\rangle =
\exp[a_1+a_2T_{9}^{-1}+a_3T_{9}^{-1/3}+a_4T_{9}^{1/3}\nonumber\\
+a_5T_{9}+a_6T_{9}^{5/3}+a_7\ln(T_{9})]\nonumber\\
+\exp[a_8+a_9T_{9}^{-1}+a_{10}T_{9}^{-1/3}+a_{11}T_{9}^{1/3}\nonumber\\
+a_{12}T_{9}+a_{13}T_{9}^{5/3}+a_{14}\ln(T_{9})].
\end{eqnarray}
The coefficients $a_i$ for the central value, lower limit and upper
limit of the $^{12}$N($p,\,\gamma$)$^{13}$O reaction rate are listed
in Table \ref{tab2}. The fitting errors are all less than 6\% in a
range from $T_{9}=0.01$ to $T_{9}=10$.

Since there is the large discrepancy between the rates in this work
and those in Ref. \cite{wie89}, the temperature and density
conditions at which the rap-processes are expected to operate need
to be revised. We performed reaction network simulations with a
series of constant temperatures (0.1-1.5 GK) and densities (1-10$^8$
g/cm$^3$), and a burning time of 100 s, and primordial abundances
and reaction rates from REACLIB as an initial input.

\begin{figure}[htbp]
\begin{center}
\resizebox{0.45\textwidth}{!}{
\includegraphics{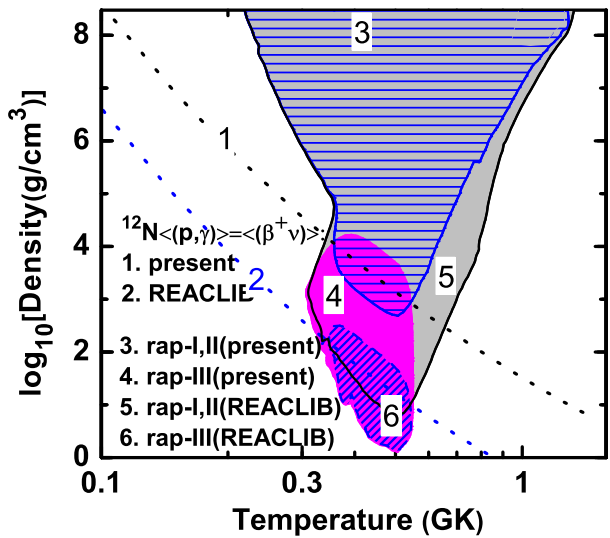}}
\caption{(Color online) Temperature and density conditions at which
the $^{12}$N($p,\,\gamma$)$^{13}$O reaction could operate. Curve 1
represents the equilibrium lines between the rates of the
$^{12}$N($p,\,\gamma$)$^{13}$O reaction and $^{12}$N $\beta^+$
decay. Curve 2 shows the same result determined from Ref.
\cite{wie89}. Regions 3 and 4 denote the revised temperature-density
conditions for rap-I,II and rap-III with the present $^{12}$N($p,\,\gamma$)$^{13}$O rate,
respectively, while Regions 5 and 6 represent those with the REACLIB
rate from Ref. \cite{wie89}. Within these four regions more
than 1$\times$10$^{-6}$ abundance (mass fraction/mass number) could
be converted to CNO cycle. Note that when determining Regions 3 and 4,
only the $^{12}$N($p,\,\gamma$)$^{13}$O rate was changed, all the rest were
still taken from the REACLIB compilations.} \label{fig:8}
\end{center}
\end{figure}

In Fig. \ref{fig:8} we shows the resulting temperature-density
conditions for the rap-I,II and rap-III processes by using the
present $^{12}$N($p,\,\gamma$)$^{13}$O rates and those of Ref.
\cite{wie89}. Curve 1 indicates the present conditions at which the
$^{12}$N($p,\,\gamma$)$^{13}$O reaction has equal strength with the
competing $\beta^+$ decay of $^{12}$N. Below this curve, the
$^{12}$N $\beta^+$ decay will prevail over its proton capture and
lead to $^{12}$C. Curve 2 shows the same result determined from Ref.
\cite{wie89}. In Regions 3-6, more than 10$^{-6}$ abundances (mass
fraction/mass number) could be converted to CNO cycle. One sees that
the present region for rap-I and rap-II (Region 3), where the
$^{12}$N($p,\,\gamma$)$^{13}$O reaction operates, was significantly
reduced relative to that from the compilation (Region 5). Therefore,
the lower limit of the density, where the 10$^{-6}$ abundance can be
converted to CNO cycle, was raised from $\sim$10 to $\sim$1000
g/cm$^3$. This is because the new rates are about two orders of
magnitude slower than the compilation. On the contrary, the present
region for rap-III (Region 4), where the $\beta$-decay of $^{12}$N
prevails over its proton capture, was enlarged relative to Region 6,
which led to an increase of the upper limit of the density from
$\sim$100 to $\sim$10000 g/cm$^3$.

In brief, the present rate of $^{12}$N($p,\,\gamma$)$^{13}$O shows
that it will only compete successfully with the $\beta^+$ decay of
$^{12}$N at higher ($\sim$two orders of magnitude) densities than
initially predicted in Ref. \cite{wie89}. This finding is consistent
with the result reported in Ref. \cite{ban09}, while is contrary to
that in Ref. \cite{sko07}.

\section{Summary and conclusion}

In this work, the angular distribution of the
$^2$H($^{12}$N,\,$^{13}$O$_\mathrm{g.s.}$)$n$ reaction was measured
and utilized to derive the ANC for the virtual decay of
$^{13}$O$_\mathrm{g.s.}$ $\rightarrow$ $^{12}$N + $p$. Our result is
in agreement with that from the
$^{14}$N($^{12}$N,\,$^{13}$O)$^{13}$C transfer reaction in Ref.
\cite{ban09}. The astrophysical S-factors and rates for the direct
capture in the $^{12}$N($p,\,\gamma$)$^{13}$O reaction were then
obtained from the measured ANC by using the direct radiative capture
model. In addition, we determined the total S-factors and reaction
rates by taking into account the direct capture into the ground
state of $^{13}$O, the resonant capture via the first excited state
of $^{13}$O and the interference between them. This work provides an
independent examination to the existing results on the
$^{12}$N($p,\,\gamma$)$^{13}$O reaction. We conclude that the direct
capture dominates the $^{12}$N($p,\,\gamma$)$^{13}$O reaction for
the temperatures below $T_{9}$ = 1.5.

We also performed reaction network simulations with the new rates.
The results imply that $^{12}$N($p,\,\gamma$)$^{13}$O will only
compete successfully with the $^{12}$N $\beta^+$ decay at higher
($\sim$two orders of magnitude) densities than initially predicted
in Ref. \cite{wie89}. Recent simulation of massive metal-free stars
between 120 and 1000 solar masses shows that a metallicity as small
as $\sim$1$\times$10$^{-9}$ is sufficient to stop the contraction
\cite{heg02}. Therefore, this revise of temperature-density
conditions may have substantial implications on the evolution of
these massive metal-free stars.

\begin{acknowledgments} We acknowledge the staff of AVF
accelerator for the smooth operation of the machine. We thank the anonymous referee for the helpful
comments. This work was supported by the National Natural Science Foundation of China under
Grant Nos. 11021504, 10875175, 10720101076, 10975193 and 11075219,
the 973 program under Grant No.
2013CB834406, KAKEHI of Japan under Grant No. 21340053.
\end{acknowledgments}

\end{CJK*}

\end{document}